\documentclass[12pt]{article}

\usepackage{amssymb,amsmath}
\usepackage{latexsym}
\usepackage[usenames]{color}
\usepackage[T2A]{fontenc} %Lat+Rus
\usepackage[cp1251]{inputenc} %Windows Cyrilic
\newtheorem{theorem}{Theorem}
\newtheorem{corollary}{Corollary}
\newtheorem{lemma}{Lemma}
\begin{document}

\begin{center}
{\bf Characteristics of  conservations laws of chiral-type systems
}
\end{center}

\begin{center}
A.V. Balandin  \\
Department of Mathematics and Mechanics \\
N.I.Lobatchevsky Nizhny Novgorod State University\\
23 Gagarin ave., 603950 Nizhny Novgorod, Russia    \\
e-mail: balandin@mm.unn.ru
\end{center}

\begin{abstract}

In this note  a new way to construct  characteristics of conservations
laws of integrable chiral-type systems is proposed. Some examples of such
characteristics are considered.

\end{abstract}



%35Q51 Soliton-like equations
%37K05 Hamiltonian structures, symmetries, variational principles,
%conservation laws

%37K10 Completely integrable systems, integrability tests, bi-Hamiltonian
%structures, hierarchies (KdV, KP, Toda, etc.)

%37K25 Relations with differential geometry

\section{Introduction}

Chiral-type systems (see, for example, \cite{Mesh}) are  systems
of partial differential equations of the form
\begin{equation} \label{EQ1} \Delta^\alpha \equiv U^{\alpha}_{xy} +
G^{\alpha}_{\beta\gamma}U^{\beta}_xU^{\gamma}_y +Q^{\alpha} = 0.
\end{equation}
Here, Greek indices $\alpha ,\beta ,\gamma$ range from 1 to $n$,
and  subscripts denote partial derivatives with respect to the
independent variables $x$ and $y$. The coefficients
$G^{\alpha}_{\beta\gamma},Q^{\alpha}$ are assumed to be smooth
functions of  variables  $U^1, U^2, ..., U^n$. The summation rule
over the repeated indices is also assumed.

If  system (\ref{EQ1}) is a system of Euler-Lagrange equations, then it is called
a nonlinear generalized sigma model.

Following \cite{Olver}, recall that the characteristic of a  conservation law $L=(L_1,L_2)$
of  system  (\ref{EQ1}) is
a set of functions $R = \{R_\alpha\}$ such that
$$Div\; L= D_xL_1+D_yL_2=  R_\alpha \Delta^\alpha ,$$
where $\Delta^\alpha $ denotes the l.h.s. of Eq. (\ref{EQ1}).

By integrable systems, we mean systems  admitting  a Lax representation, i.e.,
systems for which
there exist a matrix Lie algebra  $\mathfrak{g}$ and
 % Suppose %One can see that in this case
 the matrix $\mathfrak{g}$-valued functions $\widetilde{A},\widetilde{B}, S_\alpha$ such that
%Lax representation is of the form:
the following identity is fulfilled,
\begin{equation}\label{EQ2}
D_y\widetilde{A}-D_x\widetilde{B}+[\widetilde{A},\widetilde{B}]=S_\alpha \Delta^\alpha.
\end{equation}

 \textbf{Remark 1}.
Note that the set of such matrices $S_\alpha$ in the general case
% of a Lax representation
was investigated by M.Marvan \cite{Mar},
 and named a characteristic element of a Lax representation.

\bigskip

Since  system \eqref{EQ1} does not contain $U^\alpha_{xx},U^\alpha_{yy}$,
the Lax matrices $\widetilde{A},\widetilde{B}$  satisfy the conditions
$ \widetilde{A}=\widetilde{A}(U^\alpha_x, U^\beta),\;
\widetilde{B}=\widetilde{B}(U^\alpha_y, U^\beta).$

Choose these matrices  in the form

\begin{equation}\label{EQ3}
\widetilde{A} =A_\alpha U^\alpha_x + M,\;\;\; \widetilde{B}=B_\alpha U^\alpha_y +N,
\end{equation}
where  $ A_\alpha,B_\alpha,M,N$ are smooth functions of the variables $U^1,U^2,...,U^n$
taking values in
a matrix Lie algebra $\mathfrak{g}.$

\textbf{Remark 2}. In all cases known to the author, the Lax matrices for  systems
  \eqref{EQ1} are of  form \eqref{EQ3} (see, for example, \cite{Lez} ).

\bigskip

Substituting \eqref{EQ3} into \eqref{EQ2} and collecting the terms
at $U^\alpha_{xy},U^\alpha_y, U^\alpha_x$,
we  find that
\begin{equation}\label{EQ300}
S_\alpha=%S_\alpha(U^1,U^2,...,U^n)=
A_\alpha-B_\alpha,
\end{equation}
\begin{equation}\label{EQ5}
M_{,\alpha}=[ B_\alpha,M],
\end{equation}
\begin{equation}\label{EQ6}
 N_{,\alpha}=[ A_\alpha,N].
\end{equation}

Here and further,  partial derivatives are denoted by comma,
i.e., $P_{,\alpha}=\frac{\partial P}{\partial U^\alpha}.$

\bigskip

\textbf{Remark 3}.
Note that form  \eqref{EQ3} of the Lax representation and the form of the chiral-type
system \eqref{EQ1} require Eqs. \eqref{EQ5},\eqref{EQ6}.

\bigskip

\textbf{Remark 4}.  Note that  Lax matrices \eqref{EQ3} do not
contain a spectral parameter. However, if $M\ne 0$ or $N\ne 0$ it
can be introduced automatically by
the transformation $x\rightarrow\lambda x, y\rightarrow\frac{1}{\lambda}y.$
Now, we obtain the following  Lax representation with the spectral parameter
\begin{equation}\label{EQ400}
\widetilde{A}=A_\alpha U^\alpha_x + \lambda M, ~~
\widetilde{B}=B_\alpha U^\alpha_y +\frac{1}{\lambda}N.
\end{equation}
%In what follows, we consider only such cases; thus, there is no need to
%consider the spectral parameter.

\bigskip

Let $ f:\underbrace{\mathfrak{g}\times\mathfrak{g}\times ...
 \mathfrak{g}}_p \mapsto C $ be a symmetric p-linear ad-invariant form on  $\mathfrak{g},$ i.e.,
for all $x_1,x_2,...,x_p,y \in \mathfrak{g}$ the following identity  holds:
$$
f([y,x_1],x_2,...,x_p)+f(x_1,[y,x_2],x_3,...,x_p)+ ...+f(x_1,x_2,...,[y,x_p])=0.
$$
For $p=2$ one can take as $f$ the Killing metric of the Lie algebra $\mathfrak{g}.$

\textbf{Remark 5}. Let $G$ be a matrix Lie group with the Lie algebra $\mathfrak{g}.$
Consider a gauge transformation
$$\widetilde{A}\rightarrow T_xT^{-1}+ T\widetilde{A}T^{-1},
\;\widetilde{B}\rightarrow T_yT^{-1}+ T\widetilde{B}T^{-1}$$
generated by a matrix $T \in G.$
 Under such transformation, the functions $S_\alpha,M,N $ are transformed according to the rule:
 $$S_\alpha\rightarrow TS_\alpha T^{-1},M \rightarrow TMT^{-1},N \rightarrow TNT^{-1},$$
and the tensor fields obtained by substituting functions  $S_\alpha,M,N $ into
ad-invariant forms on $\mathfrak{g}$ are invariant.
Thus, it seems reasonable to assume that these tensor fields  carry important information about
the integrable PDE system.

\bigskip

The  purpose of this paper is to prove the following main theorem that states that
the tensor fields $R_{\alpha}$ and $\widetilde{R}_{\alpha} $ defined by the expressions
\begin{equation}\label{EQ9}
R_{\alpha}=f(S_\alpha,\underbrace{M,...,M}_{p-1})),
\end{equation}
\begin{equation}\label{EQ10}
\widetilde{R}_{\alpha}=f(S_\alpha,\underbrace{N,...,N}_{p-1}))
\end{equation}

are  characteristics of  conservation laws of  system  (\ref{EQ1}).

\medskip
\section{Main theorem and examples}

Denote
the $\alpha-$th Euler operator   by
$$E_\alpha= \sum_{J}(-D)_J(\frac{\partial}{\partial U^\alpha_J}),$$
where the sum extends over all multi-indices $J=(j_1,j_2).$

For the sequel, we need the following technical lemma 1.

\begin{lemma}
Let matrices $\widetilde{A}$ and $\widetilde{B}$ be of form \eqref{EQ3} where $A_\alpha,B_\alpha,M,N$ are smooth
functions of the variables $U^1,U^2,...,U^n$ that take values in
a matrix Lie algebra $\mathfrak{g}$  and satisfy (\ref{EQ5}),(\ref{EQ6}).
%$$
%\widetilde{A} =A_\alpha U^\alpha_x +M, \;\widetilde{B}=B_\alpha U^\alpha_y +N,
%$$
%where $ A_\alpha,B_\alpha,M,N$ are smooth functions of the variables $U^1,U^2,...,U^n$, taking values in
%a matrix Lie algebra $\mathfrak{g},$ and such that the following
%conditions are satisfied:
%\begin{equation}\label{EQ5}
%M_{,\alpha}=[ B_\alpha,M],
%\end{equation}
%\begin{equation}\label{EQ6}
% N_{,\alpha}=[ A_\alpha,N].
%\end{equation}
%Here and further, the comma denotes the partial derivatives,
%that is, $P_{,\alpha}=\frac{\partial P}{\partial U^\alpha}.$
 Assume that $ f$
 %:\underbrace{\mathfrak{g}\times\mathfrak{g}\times ...
 %\mathfrak{g}}_p \mapsto R $
 is a symmetric p-linear ad-invariant form on  $\mathfrak{g}.$ %i.e.

 Then the following identities hold: %are valid:
\begin{equation}\label{EQ7}
E_{\alpha}(f(D_y\widetilde{A}-D_x\widetilde{B}+[\widetilde{A},\widetilde{B}],\underbrace{M,...,M}_{p-1}))=0,
\end{equation}
\begin{equation}\label{EQ8}
E_{\alpha}(f(D_y\widetilde{A}-D_x\widetilde{B}+[\widetilde{A},\widetilde{B}],\underbrace{N,...,N}_{p-1}))=0.
\end{equation}

\end{lemma}

We will prove this lemma in the appendix.

\begin{theorem}
Let the chiral-type system (\ref{EQ1}) admit the Lax representation of the form (\ref{EQ2}),(\ref{EQ3})
in a matrix Lie algebra $\mathfrak{g}.$ %where $ \widetilde{A},\widetilde{B}$ are of the form
%(\ref{13a}).% and conditions  (\ref{EQ5}),(\ref{EQ6}) are fulfilled.
Then for each ad-invariant symmetric p-form $f$ of the Lie algebra $\mathfrak{g}$,
 the sets of functions (\ref{EQ9}),(\ref{EQ10})
are  characteristics of  conservation laws of the system (\ref{EQ1}).
\end{theorem}

\textbf{Proof}.

%Evidently,
%In the case under consideration, up to a constant factor,  we evidently %obviously
%have:
%\begin{equation}\label{EQ9}
%R_{\alpha}=f(S_\alpha,\underbrace{M,...,M}_{p-1})),
%\end{equation}
%\begin{equation}\label{EQ10}
%\widetilde{R}_{\alpha}=f(S_\alpha,\underbrace{N,...,N}_{p-1})).
%\end{equation}

We will consider the case of Eq. (\ref{EQ9}). The  case of Eq. (\ref{EQ10}) can be proved in a similar way.
One can easily verify that %in order to
 coefficients at the   first-order  derivatives $U^\alpha_x,U^\beta_y$ in
l.h.s. of Eq. (\ref{EQ2}) vanish whenever %it is necessary only in case
Eqs. (\ref{EQ5}),(\ref{EQ6}) are fulfilled.

Substituting (\ref{EQ2}) in the polynomial $f$ as the first argument, and $M$ as the remaining arguments, we
obtain the relations
$$
f(D_y\widetilde{A}-D_x\widetilde{B}+[\widetilde{A},\widetilde{B}],M,...,M)=R_\alpha [U^{\alpha}_{xy} +
G^{\alpha}_{\beta\gamma}U^{\beta}_xU^{\gamma}_y +Q^{\alpha}].
$$

The use of the lemma 1 completes the proof.

\medskip

Next, we discuss  invariant properties of  characteristics of conservation laws.

One can readily verify that under arbitrary non-degenerate transforma\-tions
of the variables
$U^1, U^2,..., U^n$ the functions $G^{\alpha}_{\beta\gamma}$ in (\ref{EQ1}) are transformed as
 coef\-ficients of an affine connection. Therefore, we
assume that $G^{\alpha}_{\beta\gamma}$ are  Christoffel symbols of an affine
connection in the local coordinate system $U^1, U^2,..., U^n$ of a space $V^n$. We say that this connection is associated to  system (\ref{EQ1}), and denote  covariant derivatives w.r.t. this connection by $\nabla_\alpha.$

%Further on,
Unless otherwise indicated, we consider  characteristics of  conservation laws  of
the form $R_\alpha = R_\alpha(U^1,U^2,...,U^n)$ only.

\medskip

\begin{theorem} A set $R_\alpha$ is the characteristic of a  conservation law of system (\ref{EQ1}) iff
the following conditions are satisfied:

1) $R_\alpha$ is a Killing covector field, i.e.,
 \begin{equation}\label{EQ11}
\nabla_{(\alpha} R_{\beta)} =0;
\end{equation}

2)
\begin{equation}\label{EQ12}
 R_{\alpha} Q^{\alpha}=const;
\end{equation}

3) the form
$\nabla_{\alpha} R_{\beta}dU^\alpha\wedge dU^\beta$
is closed, i.e.,
\begin{equation}\label{EQ13}
 d(\nabla_{\alpha} R_{\beta}dU^\alpha\wedge dU^\beta)=0.
\end{equation}

\end{theorem}

\textbf{Proof}.

As it is well known (see for example \cite{Olver}), a function $R_\alpha \Delta^\alpha$ is  a divergence iff
$E_\beta (R_\alpha \Delta^\alpha)=0.$

One can readily verify that
$$
E_\alpha (R_\beta \Delta^\beta)= K_{\alpha \beta}U^{\beta}_{xy}+
\widehat{L}_{\alpha \beta \gamma}U^\beta_x U^\gamma_y
+ L_{\alpha} ,
$$
where
$$
K_{\alpha \beta}=\nabla_{(\alpha }R_{\beta)},\;
L_{\alpha }=(R_{\delta}Q^\delta)_{,\alpha},
$$%
\begin{equation}
\label{EQ14}
\widehat{L}_{\alpha \beta \gamma}
=R_{\alpha, \beta \gamma}-(R_\delta G^\delta_{\alpha \gamma})_{,\beta}
- (R_\delta G^\delta_{\beta\alpha })_{,\gamma} +(R_\delta G^\delta_{\beta \gamma})_{,\alpha}.
\end{equation}

Now, one can see that  conditions  (\ref{EQ11}) and (\ref{EQ12}) are fulfilled.
Taking into account  condition (\ref{EQ11}),
Eq.  (\ref{EQ14}) results in
$$\widehat{L}_{\alpha \beta \gamma}% K_{\alpha \beta \gamma}
= (\nabla_{\gamma }R_{\alpha})_\beta+(\nabla_{\beta }R_{\gamma})_\alpha
+ (\nabla_{\alpha }R_{\beta})_\gamma .$$
This completes the proof.

\begin{corollary}
Let the chiral-type system (\ref{EQ1}) admit the Lax representation of the form (\ref{EQ2}),(\ref{EQ3})
in a compact semisimple Lie algebra $\mathfrak{g},$
where:

1) not all of the coefficients $Q^\alpha$ vanish;

2) the number $n$ of components of the system (\ref{EQ1}) and the dimension of
 the Lie algebra   $\mathfrak{g}$ satisfy the
condition:
$ n\le \dim \mathfrak{g} \le n+1;
$

3) the matrices  $S_\alpha$ are linearly independent.

Then
the system (\ref{EQ1}) admits at least one non-trivial conservation law whose characteristic $R_\alpha$ depends on $U^1,U^2,...,U^n $ and satisfies the conditions (\ref{EQ11}),(\ref{EQ13}), as well as
$ R_{\alpha} Q^{\alpha}=0.$

\end{corollary}

\textbf{Proof}.

First, using the Lax representation (\ref{EQ2}), we find that $2S_\alpha Q^\alpha =[M,N].$
In view of linearly independence of $S_\alpha$ we obtain that $M$ and $N$ are linearly independent too.
Let the 2-form $f$ be the Killing metric on the Lie algebra $\mathfrak{g}.$ Taking into
account non-degeneracy and positivity of the form $f$, as well as and considerations of the dimensionality, we conclude that at least
 one of the fields $R_\alpha=f(S_\alpha,M),
\widetilde{R}_\alpha=f(S_\alpha,N)$ is non-vanishing. Now, the use of  theorem 2
and the observation that $R_{\alpha} Q^{\alpha}= f(Q^{\alpha}S_\alpha,M)=f(\frac{[M,N]}{2},M)=0$
complete the proof.

\bigskip

\textbf{Remark 6}. In general  it seems hard to say how many nonzero characteristics
 can be constructed by the proposed method.
%but in the following example it is possible.
In the following example 1,   to each semisimple Lie algebra of rank $r$, we assign
a system for which  at least $r$  linearly independent characteristics can be
constructed using theorem 1.

\bigskip

\textbf{Example 1}: Systems associated to semisimple Lie groups \cite{Lez},\cite{Bal}.
%Let $G$ be an n-dimensional Lie group  with  local coordinates $U^\alpha,$
%Lie algebra $\mathfrak g,$
%and a basis of left invariant forms $\varphi^\alpha =S^\alpha_\beta dU^\beta .$
% Consider Lax representation
Let
$U^1,U^2,...U^n$ be local coordinates on a semisimple Lie group $G$ of rank $r,$
$\mathfrak g$ be Lie algebra of $G$,
and
$C^\alpha_{\beta\gamma}$ %be
the structure constants of the group $G$ corresponding to %determined by
a
basis of
left-invariant differential forms
$\Theta^{\alpha} = T^{\alpha}_{\beta}dU^{\beta}.$
%i.e. %where
Thus,
the following equations are fulfilled
$$
%\label{3d}
d\Theta^{\alpha}=
C^{\alpha}_{\beta \gamma} \Theta^{\gamma} \wedge
\Theta^{\beta},
$$%\end{equation}
\begin{equation}
\label{4d} T^{\alpha}_{[\mu
,\nu]}=C^{\alpha}_{\beta\gamma}T^{\beta}_{\mu}T^{\gamma}_{\nu}.
\end{equation}
%Thus,
%and the following
%Maurer-Cartan equations are fulfilled
Assume that  the matrices $\widetilde{A}=(\widetilde{A}^\alpha_{\beta}),
\widetilde{B}=(\widetilde{B}^\alpha_{\beta})$
are of the form %(see, for example, [?]Th 2.)
 $$\widetilde{A}^\alpha_{\beta}= 2C^\alpha_{\beta\gamma}(T^{\gamma}_{\lambda}U^{\lambda}_x +M^{\gamma}),\;\; \widetilde{B}^\alpha_{\beta}= 2C^\alpha_{\beta\gamma}%T^{\gamma}_{\lambda}U^{\lambda}_x +
N^{\gamma},$$
where $M^{\alpha} = const$ and $N^{\alpha}$
satisfy the equations
\begin{equation}
\label{d14}
N^\alpha_{,\gamma} = 2C^\alpha_{\beta \delta}T^{\delta}_{\gamma}N^{\beta}.
\end{equation}
%Note that,
Using Eqs. \eqref{4d}, one can find that  the system \eqref{d14} is  completely
integrable, and its  solutions are determined by the set of $n$ constants $N_0=(n_1,n_2,...,n_n)$.

Now, one can verify that these matrices form the Lax representation %\eqref{11111}
of  chiral-type system \eqref{EQ1}, where
$$G^{\sigma}_{\mu \nu} =
 \tilde{T}^{\sigma}_{\lambda}T^{\lambda}_{\mu ,\nu},\;
Q^{\sigma} = 2\tilde{T}^{\sigma}_{\lambda}C^{\lambda}_{\mu \nu}N^{\nu}M^{\mu},
$$
and $\tilde{T}$  denotes the inverse matrix of $T.$
Thus, such integrable systems are determined by two elements $M,N_0$ of the algebra
$\mathfrak g.$
Note that  characteristic elements $S_\alpha$ of the Lax representation
for these systems are the matrices $S_\lambda =||2C^\alpha_{\beta\gamma}T^{\gamma}_{\lambda} ||. $
These matrices  $S_\lambda  $ form a basis of the matrix algebra $\mathfrak g.$

Let $I(\mathfrak g)$ be the ring of $ad$-invariant polynomials on $\mathfrak g.$
The known theorem (%of Chevalley
see, for example, \cite{Go}) states that
the ring $I(\mathfrak g)$ is generated by precisely $r$ algebraically independent
homogeneous polynomials.
Denote the symmetric $ad$-invariant forms on $\mathfrak g$
obtained by the polarization of  the generating %such basic
polynomials by %of $I(\mathfrak g)$ by
$f_1,f_2,...,f_r$, of degrees $p_1,p_2,..,p_r$, respectively.

%Let $f_i$ be a $p_i$-linear form.
Fix $M\in \mathfrak g$ and consider
the linear functional $\widetilde{f}_i $ on $\mathfrak g$ defined by the equality
$$\widetilde{f}_i(X)=f_i(X,\underbrace{M,M,...,M}_{p_i-1}),\;X\in \mathfrak g. $$
Then,  by the result of Kostant \cite{Ko} (see, also, \cite{Var}) ,
the functionals $\widetilde{f}_1, \widetilde{f}_2,...,\widetilde{f}_r$ are linearly
independent iff $M$ is a regular element of the algebra $\mathfrak g.$

Now, turning back to the chiral-type system, %choose
assume that  $M$ is a regular element
of the algebra $\mathfrak g.$

Note that the characteristic elements $S_\alpha$ of the Lax representation
for these systems are the matrices $S_\lambda =||2C^\alpha_{\beta\gamma}T^{\gamma}_{\lambda} ||. $
Such matrices  $S_\lambda  $ form a basis of the matrix algebra $\mathfrak g.$
Thus, the characteristics of  conservation laws of the form $F_{i}=(F_{i\alpha})= (f_i(S_\alpha,\underbrace{M,M,...,M}_{p_i-1})),\;(i=\overline{1,r})$ are
linearly independent.

\bigskip

\textbf{Remark 7}. Note that the characteristics  constructed using theorem 1
are characteristics of order zero, i.e.,
they do not depend on  the derivatives, and are of the form
$F_\alpha=F_\alpha(U^1,U^2,...,U^n).
$ %In case of

Let a chiral-type
system be a
system
obtained from a Lagrangian of the form
$$L= h_{\alpha\beta}(U^1,...,U^n)U^\alpha_x U^\beta_y +Q(U^1,...,U^n), \; det (h_{\alpha\beta})\ne 0. $$
 For such
systems  characteristics obtained using theorem 1
correspond to  symmetries of the form
\begin{equation}
\label{DEQ14}
V= v^\alpha(U^1,U^2,...,U^n)\frac{\partial}{\partial U^\alpha}.
\end{equation}
Thus, the number of  different characteristics, in the case of  variational systems,
is bounded by the number of  symmetries of the form (\ref{DEQ14}).

Suppose that a variational system does not admit symmetries  of the form
(\ref{DEQ14});
then all of the characteristics $F_\alpha, \widetilde{F}_\alpha$  vanish.
For example, nonlinear Klein-Gordon equation, $U_{xy}=f(U)$, does not have
symmetries of the form $V=v(U)\frac{\partial}{\partial U}.$ Thus, all
characteristics %of the conservation laws
constructed
in the proposed  way will automatically vanish.

\medskip

\textbf{
Remark 8}. If $\mathfrak g$
is
$\mathfrak sl(2)$ or $\mathfrak so(3)$, then one can
construct first
order characteristics of  conservation laws similarly to theorem 1.
By a direct computation analogous to the proof of theorem 1, one can verify %it can be proved
that   functions
$$ Y_{\alpha}=f([S_\alpha,M],[S_\beta,M],M,...,M)U^\beta_x,$$
\begin{equation}
\label{1d}
\widetilde{Y}_{\alpha}=f([S_\alpha,N],[S_\beta,N],M,...,M)U^\beta_y,
\end{equation}
 where $f$ is a symmetric ad-invariant %form the Killing
 form
 on $\mathfrak g,$
are the first order characteristics
of  conservation laws.

This construction, and theorem 1, are illustrated  in
 examples 2,3,4 below.

\medskip

\textbf{Example 2}.
 Consider the Pohlmeier-Lund-Regge system \cite{LR},

 $$
\Delta^1=  U^1_{xy}+\frac{1}{\sin U^2}(U^1_xU^2_y+U^1_yU^2_x)=0,
 $$
  $$
\Delta^2= U^2_{xy}-\frac{\sin U^2}{(1+\cos U^2)^2}U^1_xU^1_y-p \sin U^2=0,
 $$
where $p$ is an arbitrary constant.
It will be convenient to write the Lax representation of PLR system in  form (\ref{EQ2}),(\ref{EQ3}),
where
$$
\widetilde{A}=\left (
\begin{array}{ccc}
0&\lambda p-\frac{\cos U^2 U^1_x }{2\cos^2 \frac{U^2}{2}}&-tg \frac{U^2}{2}U^1_x\\
-(p\lambda-\frac{\cos U^2 U^1_x }{2\cos^2 \frac{U^2}{2}})&0&U^2_x \\
tg \frac{U^2}{2}U^1_x&-U^2_x&0
\end{array}
\right ),
$$
$$
\widetilde{B}=\left (
\begin{array}{ccc}
0&-(\frac{\cos U^2}{\lambda}+\frac{U^1_y}{2\cos^2 \frac{U^2}{2}})&-\frac{\sin U^2}{\lambda}\\
\frac{\cos U^2}{\lambda}+\frac{U^1_y}{2\cos^2 \frac{U^2}{2}}&0&0\\
-\frac{\sin U^2}{\lambda}&0&0
\end{array}
\right ).
$$
Choose the following basis $B$ of the Lie algebra $\mathfrak{so(3)}$
$$
\overrightarrow{e}_1 =\left (
\begin{array}{ccc}
0&0&0\\
0&0&1\\
0&-1&0
\end{array}
\right ),\overrightarrow{e}_2=\left (
\begin{array}{ccc}
0&0&-1\\
0&0&0\\
1&0&0
\end{array}
\right ),\overrightarrow{e}_3=\left (
\begin{array}{ccc}
0&1&0\\
-1&0&0\\
0&0&0
\end{array}
\right );
$$
then  we have

\begin{eqnarray*}
D_y\widetilde{A}-D_x\widetilde{B}+[\widetilde{A},\widetilde{B}]=S_\alpha \Delta^\alpha=
  \left (
\begin{array}{ccc}
0&tg^2\frac{U^2}{2}&-tg\frac{U^2}{2}\\
-tg^2\frac{U^2}{2}&0&0\\
tg\frac{U^2}{2}&0&0
\end{array}
\right )\Delta^1
\\+\left (
\begin{array}{ccc}
0&0&0\\
0&0&1\\
0&-1&0
\end{array}
\right )\Delta^2.
\end{eqnarray*}
Thus, w. r. t.
the basis $B$,
$$
S_1=(0,tg\frac{U^2}{2},tg^2\frac{U^2}{2}),\;S_2=(1,0,0),$$
$$M=(0,0,p),\;N=(0, \sin U^2,-\cos U^2). $$

Assume that the 2-form $f$ is the Killing metric of the Lie algebra $\mathfrak{so}(3)$ which, w. r. t.  the  basis
$B$,  is $\delta_{ij}$ up to a constant factor.
Next, we obtain $$R=(f(S_1,M),f(S_2,M))=(ptg^2\frac{U^2}{2},0),$$
$$\widetilde{R}=(f(S_1,N),f(S_2,N))=(tg^2\frac{U^2}{2},0).$$
Indeed, one can verify that the 1-form $\phi= tg^2\frac{U^2}{2}(U^1_xdx-U^1_ydy)$
is the conservation law of PLR system with the characteristic $\widetilde{R}=\frac{1}{p}R$.

Let us construct a characteristic of the first order. Using expressions \eqref{1d} and assuming
that $f$ is the Killing form, we obtain
$$Y_1=p^2U^1_x\tan ^2\frac{U^2}{2},\; Y_2=p^2U^2_x,\; \widetilde{Y}_1=U^1_y\tan ^2\frac{U^2}{2},\;
 \widetilde{Y}_2=U^2_y. $$
One can verify that $Y$ and $\widetilde{Y}$ are
 the characteristics of the following
conservation laws: $p^2[(U^1_x)^2\tan ^2\frac{U^2}{2}+(U^2_x)^2]dx-2p^2\cos U^2dy $ and
$[(U^1_x)^2\tan ^2\frac{U^2}{2}+(U^2_y)^2]dy-2\cos U^2dx $, respectively.

\medskip

\textbf{Remark 9}.
It can be proved %one can readily verify
that PLR-system admits only one conservation law whose characteristic depends on $U^1,U^2.$
It would be interesting to investigate  in which cases of integrable chiral-type
systems all of the characteristics depending on
 $U^1,U^2,...,U^n$ can be found by using expression \eqref{EQ9},\eqref{EQ10}.

\medskip

\textbf{Example 3}.  Consider the 3-component system

$$U^1_{xy} + U^3_xU^1_y ctgU^3 -\frac{1}{\sin U^3}U^3_yU^2_x +
%a\frac{\sin U^1}{\sin U^3}
= 0, $$
$$
U^2_{xy} + U^3_yU^2_x ctgU^3 -\frac{1}{\sin U^3}U^3_xU^1_y
%- a\frac{\sin U^1 \cos U^3}{\sin U^3}
= 0,
$$%\end{equation}
$$U^3_{xy} + U^1_yU^2_x \sin U^3
- p\sin U^3 %-a\cos U^1\cos U^3
= 0, $$
where $p$ is an arbitrary constant.
This system
admits the Lax representation of the form (\ref{EQ2}),(\ref{EQ3}),
where \cite{Bal}:

$$
\widetilde{A}=\left (
\begin{array}{ccc}
0&i\lambda M^3&-i\lambda M^2\\
-\lambda M^3&0&i\lambda M^1 \\
i\lambda M^2&-i\lambda M^1&0
\end{array}
\right ),
$$
$$
\widetilde{B}=\left (
\begin{array}{ccc}
0&\frac{i}{\lambda}-(\cos U^3 U^1_y + U^2_y)&-b_{31}\\
-\frac{i}{\lambda}+(\cos U^3 U^1_y + U^2_y)&0&b_{23}\\
b_{31}&-b_{23}&0
\end{array}
\right ),
$$
$$M^1 = p\sin U^3\sin U^2,\;
M^2 = -p\sin U^3 \cos U^2 ,\;
M^3 = p\cos U^3, $$
$$
b_{31}=\sin U^3\cos U^2 U^1_y - \sin U^2U^3_y,\;\;
b_{23}=-\cos U^2 U^3_y - \sin U^2 \sin U^3U^1_y.
$$
Consider the same basis $B$ and the same 2-form f as in example 1. Then, we find
$S_1=(\sin U^2\sin U^3, - \cos U^2\sin U^3, \cos U^3)$,
$S_2=(0,0,1)$, $S_3=(\cos U^2, \sin U^2,0 ),$ and
$R=\{f(S_\alpha,M)\}=\{p,p\cos U^3,0\}, \widetilde{R} =\{f(S_\alpha,M)\}=\{\cos U^3,1,0\}.$
Now, one can see that
the set $R$,  up to factor $p$, is the characteristic of the conservation law $(\cos U^3U^2_x + U^1_x)dx,$
 and the set $\widetilde{R}$ is the characteristic of the conservation law
$(\cos U^3U^1_y + U^2_y)dy.$

Again, using \eqref{1d} and  assuming $f$ to be
the Killing form, one can find $Y=(0,U^2_x\sin^2 U^3, U^3_x),
\widetilde{Y}=(U^1_y\sin^2 U^3, -U^2_x,0) $ which are the
 first-order characteristics of the conservation laws
$ \frac{dx}{2}[(U^3_x)^2 + (U^2_x \sin U^3)^2] - p\cos U^3dy $
and $\frac{dy}{2}[(U^3_y)^2 + (U^1_y \sin U^3)^2]- p\cos U^3dx. $

\medskip

\textbf{Example 4}.
The Lax representation for the Sine-Gordon equation is
of the form:
$$
\widetilde{A}=\left(
\begin{array}{cc}
i\lambda&i\frac{U_x}{2}\\
i\frac{U_x}{2}&-i\lambda
\end{array}
\right ),\widetilde{B}=\frac{1}{4i\lambda} \left(
\begin{array}{cc}
\cos U&-i\sin U\\
i\sin U&-\cos U
\end{array}
\right ),
$$
then  $S=\left(
\begin{array}{cc}
0&\frac{i}{2}\\
\frac{i}{2}&0
\end{array}
\right ),\; M=\left(
\begin{array}{cc}
i&0\\
0&-i
\end{array}
\right ).\; %([S,M],[S,M])=const\ne 0,
$ %and
Now, using Eq. \eqref{1d},  one can obtain  a characteristic of the fist order
$Y=U_x.$ Similarly, one gets $\widetilde{Y}=U_y$.
One can see that  characteristics $ Y,\widetilde{Y}$ correspond to the conservation laws
$ \frac{1}{2}U_x^2dx-\cos Udy,\;\;\frac{1}{2}U_y^2dy-\cos Udx ,$
respectively.

\medskip

\textbf{Remark 10}.
Using \eqref{EQ400}, one can rewrite \eqref{EQ9},\eqref{EQ10}, up to a constant factor, in the following way
$$ R_{\alpha}=f(S_\alpha, \frac{\partial \widetilde{A}}{\partial \lambda},
\frac{\partial \widetilde{A}}{\partial \lambda},...,
\frac{\partial \widetilde{A}}{\partial \lambda}), \;\;\;\widetilde{R}_{\alpha}=f(S_\alpha, \frac{\partial \widetilde{B}}{\partial \lambda},\frac{\partial \widetilde{B}}{\partial \lambda},...,
\frac{\partial \widetilde{B}}{\partial \lambda}).
$$

It turns out that these expressions for characteristics are still valid not just for  chiral-type systems, but in some cases of  evolution equations. This is illustrated
in examples 5-8.

\bigskip

In the examples below,  we consider  $\mathfrak{sl(2)}$-valued
Lax representation,  %\cite{Gu}
%with Lie algebra $\mathfrak{sl(2)}$
 assume that $f$ is the Killing metric on $\mathfrak{sl(2)}$, and that $S_\alpha$ is a characteristic
element of the Lax representation.

\medskip

\textbf{Example 5}: Korteweg de Vries equation.
Write the Lax representation of KdV in the form \cite{Nov}:
$$D_t\widetilde{A}-D_x\widetilde{B}+[\widetilde{A},\widetilde{B}]= i(U_t-6UU_x+U_{xxx})\left(
\begin{array}{cc}
1&0\\
0&-1
\end{array}
\right ),$$
where
$$ \widetilde{A} =i\lambda \left(
\begin{array}{cc}
1&0\\
0&-1
\end{array}
\right )+i\left(
\begin{array}{cc}
0&U\\
1&0
\end{array}
\right ),$$
$$\widetilde{B}=-4\lambda^2\widetilde{A}-2i\lambda \left(
\begin{array}{cc}
-U&-iU_x\\
0&U
\end{array}
\right )+\left(
\begin{array}{cc}
U_x&iU_{xx}+2iU^2\\
2iU&-U_x
\end{array}
\right ).$$
Then, we find $S=i\left(
\begin{array}{cc}
0&1\\
0&0
\end{array}
\right ),$ $f(S, \frac{\partial \widetilde{A} }{\partial \lambda})=0$,
$f(S,\frac{\partial \widetilde{B}}{\partial \lambda})=8i\lambda $ and $f(S,\frac{\partial \widetilde{B}}{\partial \lambda}) (U_t-6UU_x+U_{xxx})=D_t(U)+D_x(-3U_x^2+U_{xx}).$

\medskip

\textbf{Example 6}: Zakharov-Shabat, or Ablowitz-Kaup-Newell-Segur system has the form
$$\triangle^1=U_t-U_{xx}-2U^2V=0,\; \triangle^2= V_t+V_{xx}+2UV^2=0.$$
ZS-AKNS system has the %$\mathfrak{sl(2)}$-valued
Lax representation %taking value in $\mathfrak{sl(2)}$
of the form \cite{Adl}:
$$ \widetilde{A} = \left(
\begin{array}{cc}
\lambda&-V\\
U&-\lambda
\end{array}
\right ),\widetilde{B}=-2\lambda\widetilde{A} - \left(
\begin{array}{cc}
-UV&V_x\\
U_x&UV
\end{array}
\right ),$$
$$D_t\widetilde{A}-D_x\widetilde{B}+[\widetilde{A},\widetilde{B}]= \left(
\begin{array}{cc}
0&0\\
1&0
\end{array}
\right )\triangle^1+ \left(
\begin{array}{cc}
0&-1\\
0&0
\end{array}
\right )\triangle^2. $$
%Thus, in the case under consideration,
%Assume $f$ again to be the Killing metric of the Lie algebra $\mathfrak{sl(2)},$
Then,
we obtain $$ f(S_\alpha, \frac{\partial \widetilde{A} }{\partial \lambda})=0 $$ and
 $$ f(S_1, \frac{\partial \widetilde{B} }{\partial \lambda}) =-2V,\; f(S_2, \frac{\partial \widetilde{B} }{\partial \lambda})=-2U.$$ Now,  the following identity holds:
$$-2V\triangle^1-2U\triangle^2=-D_t(2UV)-D_x(UV_x-VU_x). $$

\medskip

\textbf{Example 7}: AKNS $2\times2$ system of the third order \cite{Gu}.
Let us consider the system
$$\Delta_1=-U_t-\frac{1}{4}\alpha_0U_{xxx}-\frac{1}{2}\alpha_1U_{xx}+\frac{3}{2}\alpha_0U_{x}UV
-\alpha_2U_{x}+\alpha_1U^2V-2\alpha_3U=0,$$
$$\Delta_2=-V_t-\frac{1}{4}\alpha_0V_{xxx}+\frac{1}{2}\alpha_1V_{xx}+\frac{3}{2}\alpha_0V_{x}UV
-\alpha_2V_{x}+\alpha_1UV^2+2\alpha_3V=0,
$$
where $\alpha_i\; (i=\overline{0,3}) $ are arbitrary constants.
This system admits the Lax matrices of the form
$$ \widetilde{A} = \left(
\begin{array}{cc}
\lambda&U\\
V&-\lambda
\end{array}
\right ),\;
\widetilde{B}= \left(
\begin{array}{cc}
\widetilde{b}_{11}&\widetilde{b}_{12}\\
\widetilde{b}_{21}&-\widetilde{b}_{11}
\end{array}
\right ).$$
Here $$ \widetilde{b}_{11}= \alpha_0\lambda^3+  \alpha_1\lambda^2 +\lambda(\alpha_2-\frac{1}{2}UV)
+\frac{1}{4}(UV_x-VU_x)-\frac{1}{2}\alpha_1UV+\alpha_3,$$
$$ \widetilde{b}_{12}= \alpha_1U\lambda^2+\lambda(\frac{1}{2}\alpha_0U_x+\alpha_1U)+
\frac{1}{4}\alpha_0(U_{xx}-2U^2V)+\frac{1}{2}\alpha_1U_{x}+\alpha_2U,
$$
$$\widetilde{b}_{21}=\alpha_1V\lambda^2+\lambda(-\frac{1}{2}\alpha_0V_x+\alpha_1V)+
\frac{1}{4}\alpha_0(V_{xx}-2UV)-\frac{1}{2}\alpha_1V_{x}+\alpha_2V.$$
Then, we obtain $f(S_\alpha, \frac{\partial \widetilde{A} }{\partial \lambda})=0$, and the equality
$f(S_\alpha, \frac{\partial \widetilde{B} }{\partial \lambda})\Delta^\alpha =\alpha_0\lambda(V \Delta^1+U\Delta^2)+
(\alpha_1V-\frac{1}{2}\alpha_0V_x)\Delta^1+(\alpha_1U+\frac{1}{2}\alpha_0U_x)\Delta^2.$
Now, one can verify that
$$(\alpha_0U,\alpha_0V); %(-V_x,U_x),
(\alpha_1V-\frac{1}{2}\alpha_0V_x,\;\alpha_1U+\frac{1}{2}\alpha_0U_x) $$
are the characteristics of conservation laws.

\medskip

\textbf{
Example 8}: AKNS $2\times2$ system of the forth  order \cite{Gu}.
Here we consider the system
$$\Delta_1=-U_t-U_{xxxx}+8UVU_{xx}+6U^2_xV+4UU_xV_x+2U^2V_{xx}-6U^3V^2=0,$$
$$\Delta_2=-V_t+V_{xxxx}-8UVV_{xx}-6V^2_xU-4VU_xV_x-2V^2U_{xx}+6U^3V^2=0.
$$
The Lax matrices for this system are:
$$ \widetilde{A} = \left(
\begin{array}{cc}
\lambda&U\\
V&-\lambda
\end{array}
\right ),\;
\widetilde{B}= \left(
\begin{array}{cc}
\widetilde{b}_{11}&\widetilde{b}_{12}\\
\widetilde{b}_{21}&-\widetilde{b}_{11}
\end{array}
\right ),$$
where $$ \widetilde{b}_{11}= -8\lambda^4+4UV\lambda^2+2\lambda(VU_x-UV_x)+UV_{xx}+VU_{xx}-U_xV_x-3U^2V^2,$$
$$ \widetilde{b}_{12}=-8U\lambda^3-4\lambda^2U_x -2\lambda U_{xx}+ 4U^2V\lambda -U_{xxx}+6UU_{x}V,$$
$$\widetilde{b}_{21}=-8V\lambda^3+4\lambda^2V_x-2\lambda V_{xx}+ 4UV^2\lambda +V_{xxx}-6VV_xU.$$
In this case again $f(S_\alpha, \frac{\partial \widetilde{A} }{\partial \lambda})=0.$
Collecting  terms at different powers of $\lambda$  in the expressions % equality
$f(S_\alpha,\frac{\partial\widetilde{B}}{\partial \lambda}),$ similarly to example 7, we find  the
following three characteristics of conservation laws
$$(V,U);\;(V_x,U_x);\;(-2V_{xx}+4UV^2,\;-2U_{xx}+4U^2V).$$ %are the characteristics of conservation laws.

\medskip

\textbf{Example 9}.  The following evolution equation
(which is nothing but MKdV up to a transformation of $t$),
$$\triangle=-U_t+t(U_{xxx}+6U^2U_x) =0,$$
admits the Lax matrices \cite{Gu}
$$
\widetilde{A} = \left(
\begin{array}{cc}
\lambda&U\\
-U&-\lambda
\end{array}
\right ),\;
\widetilde{B}= \left(
\begin{array}{cc}
\widetilde{b}_{11}&\widetilde{b}_{12}\\
\widetilde{b}_{21}&-\widetilde{b}_{11}
\end{array}
\right ),
$$
where
$$\widetilde{b}_{11}= -4\lambda^3t-2\lambda tU^2,\; \widetilde{b}_{12}=
 -4\lambda^2 tU-2\lambda tU_x-t(U_{xx}+2U^3),
 $$
$$\widetilde{b}_{21}= 4\lambda^2 tU-2\lambda tU_x+t(U_{xx}+2U^3).
$$
One can find that $f(S,\frac{\partial \widetilde{B}}{\partial \lambda})=16tU\lambda.$
But in this case the  equality
$E_U(tU\triangle)=0$  fails. This example shows that formula
$f(S,\frac{\partial \widetilde{B}}{\partial \lambda}) $ does not always give  characteristics of
 conservation laws.

\section{Conclusion}

In this paper we prove that expressions \eqref{EQ9},\eqref{EQ10} are  characteristics of  conservations laws for  chiral-type systems admitting a Lax representation.
According to  the examples shown above we can propose the following  conjecture. Assume that an
evolutionary AKNS system,  whose coefficients do not depend on $t$, admits a
$\mathfrak g$-valued Lax representation. Note that by construction of \cite{Gu}, coefficients
of such systems do not depend on $x.$ Then
the expressions
$$R_{\alpha}=f(S_\alpha, \frac{\partial \widetilde{A}}{\partial \lambda}),\;
\widetilde{R}_{\alpha}=f(S_\alpha, \frac{\partial \widetilde{B}}{\partial \lambda}),$$
where $f$ is the Killing
metric on the Lie algebra $\mathfrak g$, and $S_\alpha$ is the characteristic element of
the Lax representation, form the characteristic of a conservation laws of the system under
consideration.

\section{Acknowledgment}

The author is grateful to E.V. Ferapontov for
attention to this work and to Y.V. Tuzov, M.I. Kuznetsov, O.N. Kashcheeva, and E.M. Makarov for
discussions. The author is thankful to the referees for useful remarks.

This work was  supported by the project 1014 of Russian Ministry of Education and Science.

\newpage
\section{Appendix}

Here, we give proof of lemma 1.

Let us first prove
Eq. (\ref{EQ7}). The case of Eq. (\ref{EQ8}) can be proved in a similar way.
Denote by $R_{1\alpha}=E_{\alpha}(f(D_y\widetilde{A},\underbrace{M,...,M}_{p-1})), R_{2\alpha}
=E_{\alpha}(D_x\widetilde{B},\underbrace{M,...,M}_{p-1})),
R_{3\alpha} =
E_{\alpha}(f([\widetilde{A},\widetilde{B}]),\underbrace{M,...,M}_{p-1})).$ It is readily  verified that collecting  the terms
at $U^\beta_{xy},U^\beta_{x}U^\gamma_{y}, U^\beta_{x} $   in  % Eq.(\ref{13})
$R_{i\alpha},i=1,2,3,$
taking into account Eqs.(\ref{EQ5}), (\ref{EQ6}), and ad-invariancy of the form $f,$
we obtain the relations
$$
R_{i\alpha}=Z_{i\alpha\beta}U^\beta_{xy}+T_{i\alpha\beta \gamma}U^\beta_{x}U^\gamma_{y}+
W_{i\alpha\beta}U^\beta_{x}  ,
$$
where
\begin{equation}\label{EQ18}
Z_{1\alpha\beta}=2(p-1)f(A_{(\alpha},M_{,\beta)},M,...,M),
\end{equation}
$$Z_{2\alpha\beta}=2(p-1)f(B_{(\alpha},M_{,\beta)},M,...,M) $$
$$=2(p-1)f(B_{(\alpha},[B_{,\beta)},M],M,...,M)=0,$$
and
\begin{equation}\label{EQ16}
Z_{3\alpha\beta}=-2(p-1)f([A_{(\alpha},B_{\beta)}] ,M,...,M),
\end{equation}

\begin{eqnarray}\nonumber
T_{1\alpha\beta \gamma}=(p-1)\{2f(A_{[\alpha,\beta]},M_{,\gamma},M,...,M)+
f(A_{\beta,\gamma},M_{,\alpha},M,...,M)
\label{EQ17}\\
+ f(A_\alpha,M_{,\gamma\beta},M,...,M)+
(p-2)f(A_\alpha,M_{,\gamma},M_{,\beta},\underbrace{M,...,M}_{p-3})\},
\end{eqnarray}

\begin{eqnarray}\nonumber
T_{2\alpha\beta \gamma}=(p-1)\{2f(B_{[\alpha,\gamma]},M_{,\beta},M,...,M)+
f(B_{\gamma,\beta},M_{,\alpha},M,...,M)
\label{EQ19}\\
+ f(B_\alpha,M_{,\gamma\beta},M,...,M)+
(p-2)f(B_\alpha,M_{,\gamma},M_{,\beta},\underbrace{M,...,M}_{p-3})\},
\end{eqnarray}

\begin{eqnarray}\nonumber
T_{3\alpha\beta \gamma}=f([A_{\beta},B_\gamma]_{,\alpha},M,...,M)- f([A_{\alpha},B_\gamma]_{,\beta},M,...,M)
\\ \nonumber
-f([A_{\beta},B_\alpha]_{,\gamma},M,...,M)
+(p-1)\{f([A_{\beta},B_\gamma],M_{,\alpha}M,...,M)
\label{EQ20}\\
- f([A_{\alpha},B_\gamma],M_{,\beta},M,...,M)-f([A_{\beta},B_\alpha],M_{,\gamma},M,...,M)\},
\end{eqnarray}

\begin{eqnarray}\nonumber
W_{1\alpha\beta}=0,\; W_{2\alpha\beta}=(p-1)(
f(N_{,\beta},M_{,\alpha},M,...,M)
\label{EQ2OA}\\
- f(N_{,\alpha},M_{,\beta},M,...,M)),
\end{eqnarray}

\begin{eqnarray}\nonumber
 W_{3\alpha\beta}=f([A_{\beta},N]_{,\alpha},M,...,M)- f([A_{\alpha},N]_{,\beta},M,...,M)
\label{EQ21}\\
+(p-1)(f([A_{\beta},N],M_{,\alpha},M,...,M)- f([A_{\alpha},N],M_{,\beta},M,...,M)).
\end{eqnarray}

We claim that all of the coefficients of the expression $R_{1\alpha}-R_{2\alpha}+
R_{3\alpha} $ vanish.

In view of relations (\ref{EQ18}),(\ref{EQ16}), and (\ref{EQ5}),
we obtain the condition
$$
 Z_{1\alpha\beta}-Z_{2\alpha\beta}+Z_{3\alpha\beta}=0.
$$
Further, using Eqs. (\ref{EQ17})-(\ref{EQ20}) and the identities of the form
$$f([A_{\beta},B_\gamma]_{,\alpha},M,...,M)= f([A_{\beta},B_{\gamma , \alpha}],M,...,M) +
(p-1)f(A_{\beta, \alpha},M_\gamma,...,M),$$
we obtain the equations:
\begin{multline*}
T_{1\alpha\beta \gamma} -T_{2\alpha\beta \gamma}+T_{3\alpha\beta \gamma}=
(p-1)\{\underline{ f(A_\alpha,M_{,\gamma\beta},M,...,M)}
\\
+\underline{(p-2)f(A_\alpha,M_{,\gamma},M_{,\beta},\underbrace{M,...,M}_{p-3})}
-2f(B_{[\alpha,\gamma]},M_{,\beta},M,...,M)
\\
-f(B_{\gamma,\beta},M_{,\alpha},M,...,M)-f(B_\alpha,M_{,\gamma\beta},M,...,M)
\\-(p-2)f(B_\alpha,M_{,\gamma},M_{,\beta},\underbrace{M,...,M}_{p-3} )
+f([A_{\beta},B_\gamma],M_{,\alpha}M,...,M)
\\
- \underline{f([A_{\alpha},B_\gamma],M_{,\beta},M,...,M)}
-f([A_{\beta},B_\alpha],M_{,\gamma},M,...,M)\}
\\
+f([A_{\beta},B_{\gamma , \alpha}],M,...,M)- \underline{f([A_{\alpha},B_{\gamma,\beta}],M,...,M)}
-f([A_{\beta},B_{\alpha,\gamma}],M,...,M).
\end{multline*}

Denote  the sum of the underlined terms by $D_{\alpha\beta\gamma}$. We claim that $D_{\alpha\beta\gamma}=0.$ Indeed, using the identities
$$
 f([A_{\alpha},B_{\gamma,\beta}],M,...,M)=(p-1)f(A_{\alpha},[B_{\gamma,\beta},M],M,...,M),
$$

\begin{eqnarray}\nonumber
 f([A_{\alpha},B_\gamma],M_\beta,M,...,M)=f(A_{\alpha},[B_\gamma,M_\beta],M,...,M)
\label{EQ22}\\
+(p-2)f(A_{\alpha},M_\gamma,M_\beta,\underbrace{M,...,M}_{p-3}),
\end{eqnarray}
rewrite $D_{\alpha\beta\gamma}$
in the form
\begin{multline*}
D_{\alpha\beta\gamma}=(p-1)[f(A_\alpha,M_{,\gamma\beta},M,...,M)+
 (p-2)f(A_\alpha,M_{,\gamma},M_{,\beta},\underbrace{M,...,M}_{p-3} )\\
- f(A_{\alpha},[B_\gamma,M_\beta],M,...,M)-
(p-2)f(A_{\alpha},M_\gamma,M_\beta,\underbrace{M,...,M}_{p-3})\\
- f(A_{\alpha},[B_{\gamma,\beta},M],M,...,M)].
\end{multline*}
Thus we conclude that $D_{\alpha\beta\gamma}=0.$

Define $D_{1\alpha\beta\gamma}$ as follows:
\begin{eqnarray*}
D_{1\alpha\beta\gamma} = (p-1)[f([A_{\beta},B_\gamma],M_{,\alpha}M,...,M)
-f([A_{\beta},B_\alpha],M_{,\gamma},M,...,M)]\\
 +f([A_{\beta},B_{\gamma , \alpha}],M,...,M)
-f([A_{\beta},B_{\alpha,\gamma}],M,...,M).
\end{eqnarray*}

Using Eq. (\ref{EQ22}), one can verify that $D_{1\alpha\beta\gamma}$
results in
$$ D_{1\alpha\beta\gamma}=(p-1)f(A_{\beta},M_{,\gamma\alpha}-M_{,\alpha\gamma},M,...,M),$$
i.e., $ D_{1\alpha\beta\gamma} =0.$

 Denote by
 $D_{2\alpha\beta\gamma}=T_{1\alpha\beta \gamma} -T_{2\alpha\beta \gamma}+T_{3\alpha\beta \gamma}.$
 Now taking into account the  previous computations, $D_{2\alpha\beta\gamma}$ yields that
\begin{eqnarray*}
 D_{2\alpha\beta\gamma}=-(p-1)\{
f(B_{\alpha,\gamma},M_{,\beta},M,...,M)-f(B_{\gamma,\alpha},M_{,\beta},M,...,M)\\
+f(B_{\gamma,\beta},M_{,\alpha},M,...,M)
+f(B_\alpha,M_{,\beta\gamma},M,...,M)
\\
+(p-2)f(B_\alpha,M_{,\gamma},M_{,\beta},\underbrace{M,...,M}_{p-3} )\}.
\end{eqnarray*}

Further, using the following identities:
\begin{eqnarray*}
f(B_\alpha,M_{,\beta\gamma},M,...,M)=  f(B_\alpha,[B_{\beta,\gamma },M],M,...,M)\\
+f(B_\alpha,[B_\gamma, M_{,\beta}],M,...,M)=
 -f(M{,_\alpha},B_{\beta,\gamma },M,...,M)\\
-f([B_\beta,B_\alpha], M_{,\gamma},M,...,M)-
(p-2)f(B_\alpha, M_{,\gamma},M_{,\beta},M,...,M),
\end{eqnarray*}
 $D_{2\alpha\beta\gamma}$ results in
\begin{eqnarray*}
D_{2\alpha\beta\gamma}= -(p-1)\{2f(B_{[\gamma ,\beta]},M_{,\alpha},M,...,M)+
2f(B_{[\alpha,\gamma ]},M_{,\beta},M,...,M)\\
-f([B_\beta,B_\alpha], M_{,\gamma},M,...,M).
\end{eqnarray*}
Taking into account the identities:
$$
 2[B_{[\alpha,\beta]},M]= -[B_{\alpha}, M_{,\beta}]+ [B_{\beta}, M_{,\alpha}]=-[[B_{\alpha},B_{\beta}], M],
$$
\begin{eqnarray*}
2f(B_{[\gamma ,\beta]},M_{,\alpha},M,...,M)= 2f(B_{[\gamma ,\beta]},[B_\alpha,M],M,...,M) =\\
-2f([B_{[\gamma ,\beta]},M],B_\alpha,M,...,M)
=f([[B_{\gamma},B_{\beta}], M]],B_\alpha,M,...,M)=\\
-f([B_{\gamma},B_{\beta}], M_\alpha,M,...,M),
\end{eqnarray*}

transform $D_{2\alpha\beta\gamma}$ to the form:

\begin{eqnarray*}
D_{2\alpha\beta\gamma}=(p-1)\{f([B_{\gamma},B_{\beta}], M_{,\alpha},M,...,M)+
f([B_{\alpha},B_{\gamma}], M_{,\beta},M,...,M)\\
+f([B_\beta,B_\alpha], M_{,\gamma},M,...,M)\}.
\end{eqnarray*}

In view of relations
$$ f([B_\alpha[B_{\gamma},B_{\beta}]], M,...,M)=-(p-1)f([B_{\gamma},B_{\beta}], M_{,\alpha},M,...,M)$$ and Jacobi's identity, $ D_{2\alpha\beta\gamma}$ vanishes.

Finally, denote by
\begin{equation}\label{EQ23}
 \triangle_{1\alpha\beta}=W_{1\alpha\beta}-W_{2\alpha\beta}+W_{3\alpha\beta}.
\end{equation}
 Substituting Eq. (\ref{EQ2OA}) and (\ref{EQ21}) in (\ref{EQ23}), we obtain the relations
$$\triangle_{1\alpha\beta}=f([A_{\beta},N]_{,\alpha},M,...,M)- f([A_{\alpha},N]_{,\beta},M,...,M).$$
This equation results in $$\triangle_{1\alpha\beta}= f(N_{,\alpha\beta}-N_{,\beta\alpha},M,...,M)=0.$$

\bibliographystyle{amsplain}

\begin{thebibliography}{15}

\bibitem{Bal} Balandin A.V., Pakhareva O.N., Potemin G.V.:
Lax representation of the chiral-type field equations, Phys. Lett. A. 23
168 -176 (2001).



\bibitem{Mesh}
Demskoi D.K., Meshkov A.G.:  Zero-curvature representation for a
chiral-type three-field system, Inverse Problems.  19 563--571 (2003)

\bibitem{Go} Goto M., Grosshans F.D.: Semisimple Lie algebras.
Lecture Notes in Pure and Applied Mathematics. Vol 38, (1978)

\bibitem{Gu} Gu C., Hu H., Zhou Z.: Darboux Transformations
in Integrable Systems.
Theory and their Applications to Geometry,
Springer (2005)

\bibitem{Ko}  Kostant, В.: Lie group representations on polynomial rings,
Amer. J. Math. 85 327-404 (1963)




\bibitem{Lez} Leznov A.N., Savel'ev M.V.: Group-Theoretical Methods
for Integration of
Nonlinear Dynamical Systems,
Birkhauser, (1992)


\bibitem{LR} Lund F., Regge T.:   Unified approach to strings and vortices with soliton
solutions, Phys.Rew. D. 14 1524–1535 (1976)

\bibitem{Mar} Marvan M.: On zero-curvature representations of partial differential equations.
 in: Kowalski O., Krupka D. (eds.) 5th International Conference on Differential Geometry and
Its Applications.  pp. 103-122, Opava, Czech Republic (1993),
(http://www.emis.de/proceedings/5ICDGA).



\bibitem{Nov} Novokshenov V.Y.:  Vvedenie v teoriju solitonov, IKI, Izhevsk (2002)
(in Russian)

\bibitem{Olver}  Olver P.: Applications of Lie Groups to Differential Equations, 2nd ed.
Springer, New York (1993)


\bibitem{Adl} Shabat A.B.(ed): Encyclopedia of integrable systems, version 0043,
 L.D. Landau Institute for Theoretical Physics, Moscow (2010)

\bibitem{Var} Varadarajan V.S.: On the ring of invariant polynomials on a semisimple Lie algebra,
Amer. J. Math. 90 308-317 (1968)



\end{thebibliography}

\end{document}